\documentclass[12pt,article]{iopart}

\usepackage{epsfig}
\usepackage{amscd,amssymb}
\usepackage{graphicx,xspace}
\usepackage{pstricks}
\usepackage[left=3cm,top=3cm,right=3cm,bottom=3cm]{geometry}
\usepackage{color}

\begin{document}
\newcommand \be  {\begin{equation}}
\newcommand \bea {\begin{eqnarray} \nonumber }
\newcommand \ee  {\end{equation}}
\newcommand \eea {\end{eqnarray}}

\title[High values and multifrcatal patterns]
{Counting function fluctuations and extreme value threshold in multifractal patterns: the case study of an ideal $1/f$ noise\footnote{Published: {\bf Journal of Statistical Physics: Volume 149, Issue 5 (2012), Page 898-920 }}}

\author{Yan V Fyodorov}

\address{School of Mathematical Sciences, Queen Mary University of London\\  London E1 4NS, United Kingdom}

\author{Pierre Le Doussal}

\address{CNRS-Laboratoire de Physique Th\'eorique de l'Ecole Normale Sup\'erieure\\
24 rue Lhomond, 75231 Paris
Cedex-France\thanks{LPTENS is a Unit\'e Propre du C.N.R.S.
associ\'ee \`a l'Ecole Normale Sup\'erieure et \`a l'Universit\'e Paris Sud}}

\author{Alberto Rosso }

\address{Laboratoire de Physique Th\'eorique et Mod\`eles Statistiques, CNRS (UMR 8626) \\
Universit\'e Paris-Sud, B\^at. 100, 91405 Orsay Cedex, France }


\begin{abstract}
 Motivated by the general problem of studying sample-to-sample fluctuations in  disorder-generated multifractal patterns
we attempt to investigate analytically as well as numerically the statistics of high values of the simplest model - the ideal periodic $1/f$ Gaussian noise. Our main object of interest is the number of points ${\cal N}_M(x)$
above a level $\frac{x}{2}V_m$, with $V_m=2\ln{M}$ standing for the leading-order typical value of
the absolute maximum for the sample of $M$ points. By employing the thermodynamic formalism
we predict the characteristic scale and the precise scaling form of the distribution of ${\cal N}_M(x)$
for $0<x<2$. We demonstrate that the powerlaw forward tail of the probability density, with exponent controlled by the level $x$, results
in an important difference between the mean and the typical values of ${\cal N}_M(x)$. This can be further used to determine the typical threshold $x_m$ of extreme values in the pattern which turns out to be given by $x_m^{(typ)}=2-c\ln{\ln{M}}/\ln{M}$ with $ c=\frac{3}{2}$. Such observation provides a rather compelling explanation of the mechanism behind universality of $c$. Revealed mechanisms are conjectured to retain their qualitative validity for a broad class of disorder-generated multifractal fields. In particular, we predict that the typical value of the maximum $p_{max}$ of intensity is to be given by $-\ln{p_{max}}=\alpha_{-}\ln{M}+\frac{3}{2f'(\alpha_{-})}\ln{\ln{M}}+O(1)$, where $f(\alpha)$ is the corresponding singularity spectrum positive for $\alpha\in{\alpha_{-},\alpha_{+}}$ and 
vanishing at $\alpha=\alpha_{-}>0$. For the $1/f$ noise case we further study asymptotic values of the prefactors in scaling laws for the moments of the counting function. Our numerics shows however that one needs prohibitively large sample sizes to reach such asymptotics even with a moderate precision. 

\end{abstract}

\maketitle

\section{Introduction}

Investigations of  multifractal structures of diverse origin is for several decades a
very active field of research in various branches of applied mathematical sciences
like chaos theory, geophysics and oceanology  \cite{multif1,multif2} as well as climate studies \cite{rainfall},
 mathematical finance \cite{multifinance1,multifinance2},
 and in such areas of physics as turbulence \cite{turb,Schertzer}, growth processes \cite{DLA}, and theory of quantum disordered systems
\cite{EM}. The main characteristics of  a multifractal pattern of data is to possess high variability over
a wide range of space or time scales, associated with huge fluctuations in intensity which can be visually
detected.

To set the notations, consider a
certain (e.g. hypercubic) lattice of linear extent $L$ and lattice spacing $a$ in
$d-$dimensional space, with $M\sim (L/a)^d\gg 1$ standing for the total
number of sites in the lattice. The multifractal patterns are
then usually associated with a set of non-negative "heights" $ h_i\ge 0$ attributed to every
lattice site $i=1,2,\ldots, M$ such that the heights scale in the limit $M\to \infty$
differently at different sites: $h_i\sim M^{x_i}$
\footnote{Usually one defines the exponents $\gamma_i$ via the relation $h_i\sim
L^{\gamma_i}$ i.e. by the reference to linear scale $L$ instead of
the total number of sites $M\sim (L/a)^d$, and similarly for the density of exponents $\rho(\gamma)\sim L^{f(\gamma)}$. We however find it more
convenient to use instead the exponents $x_i=\gamma_i/d$ and the singularity spectrum $f(x)=\frac{1}{d}f(\gamma)$.}, with exponents $x_i$ forming a dense set. To characterize such a pattern of heights quantitatively it is natural to count the sites with the same scaling behaviour.
Then a multifractal measure is characterized by a (usually, concave) single-smooth-maximum {\it singularity spectrum} function $f(x)$. Denoting the position of its maximum as $x=x_0$, such function describes  the (large-deviation) scaling of the number of points in the pattern whose local exponents $x_i$ belong to some interval around $x_0$. More precisely, defining the density of exponents by $\rho_M(x)=\sum_{i=1}^M\,\delta\left(\frac{\ln{h_i}}{\ln{M}}-x\right)$  a nontrivial multifractality implies that such density should behave in the large-$M$ limit as \cite{MenSri}
\be\label{multiansatz}
\rho_M(x)\approx c_M(x) \sqrt{\ln{M}}\, M^{f(x)}\,
\ee
 with a prefactor $c_M(x)$ of the order of unity which may still depend on $x$.
 We will refer below to the above form as the {\it multifractal ansatz}. The major effort in the last decades was directed towards determining the shape and properties of $f(x)$. In contrast, our main object of interest will be the behaviour of the prefactor $c_M(x)$ which is much less
 studied, to the best of our knowledge. In particular, if the multifractal pattern is randomly generated like e.g. those considered in \cite{EM} the ansatz (\ref{multiansatz}) is expected to be valid in every realization of the disorder.  One may then be interested in understanding the sample-to-sample fluctuations of the prefactor $c_M(x)$.

 To that end we find it convenient to introduce the counting functions
 \be\label{eq1}
 N_>(x)=\int_x^{\infty}\rho_M(y)\,dy, \quad N_<(x)=\int_{-\infty}^x\rho_M(y)\,dy
 \ee
for the total number $N_>(x)$ of sites of the lattice where heights satisfy $h_i>M^x$ (respectively, $h_i<M^x$). Substituting the multifractal form of the density  into (\ref{eq1}) and performing at $\ln{M}\gg 1$ the resulting integral for $x>x_0$ by the Laplace method we find $ N_>(x)\approx c_M(x) M^{f(x)}/|f'(x)|\sqrt{\ln{M}}$
and a similar expression for $N_<(x)$ for $x<x_0$, relating  the singularity spectrum $f(x)$ to the counting functions.
 As both $N_>(x)$ and $N_{<}(x)$ can not be smaller than unity we necessarily have $f(x)\ge 0$ for all $x$, and the condition
$f(x)= 0$ defines generically the maximal $x_{+}$ and
the minimal $x_{-}$ threshold values of the exponents which can be
observed in a given height pattern.

The singularity spectrum $f(x)$ is not a quantity which is easily calculated analytically or even numerically for a given  multifractal pattern of heights \cite{MenSri}. An alternative procedure of analysing the multifractality  is frequently referred to in the literature as the {\it thermodynamic formalism} \cite{multif1,multif2}. In that approach one characterizes the multifractal pattern by the set of exponents $\zeta_q$
describing the large-$M$ scaling behaviour of the so-called {\it partition functions} $Z_q$  as
\begin{equation}\label{1}
 Z_q=\sum_{i=1}^M\, h_i^q\sim M^{\zeta_q}, \quad \ln{M}\gg 1
\end{equation}
To relate $\zeta_q$ to the singularity spectrum $f(x)$ discussed above one rewrites
(\ref{1}) in terms of the density as $Z_q=\int_{-\infty}^{\infty} M^{qx} \rho_M(x)\, dx$, and
 again employs the multifractal ansatz  (\ref{multiansatz}) for $\rho_M(x)$. Evaluating the integral in
the $\ln{M}\gg 1$ limit by the steepest descent (Laplace) method gives
\begin{equation}\label{1a}
\fl Z_q\sim c_M(x_*)  \left(\frac{2\pi}{|f''(x_*)|}\right)^{1/2}\, M^{\zeta_q} \quad \mbox{where}
\quad f'(x_*)=-q \quad \mbox{and} \quad \zeta_q=f(x_*)+q\,x_*\,,
\end{equation}
where we have assumed that $x_-<x_*<x_+$ for simplicity. This shows that the relation between $\zeta_q$ and $f(x)$ is given essentially by the {\it  Legendre transform}.
We thus see that formally the original definition of multifractality based on the density (or, equivalently, the counting functions $N_{>,<}(x)$) and the thermodynamic formalism approach (\ref{1})-(\ref{1a}) should have exactly the same content for $\ln{M}\to \infty$, provided the singularity spectrum is concave\footnote{ Examples of non-concave multifractality spectrum and the associated thermodynamic formalism
are discussed in \cite{TB}}.
Note also the normalization
identity $Z_0=\int_{-\infty}^{\infty}\rho_M(y)\,dy\equiv M$ implying $\zeta_0=1$. It also shows
that at the point of maximum $x=x_0$ we must necessarily have $f(x_0)=1$ and that $c_M(x_0)$ is indeed of the order of unity.

The formalism described above is valid for general multifractal patterns, and is insensitive to spatial organization of intensity in
the pattern. In the present paper we will be mostly interested in disorder-generated multifractal fields
whose common feature is presence of certain long-ranged powerlaw-type correlations in data values \cite{DL}. In practice, to extract singularity spectra from  a given multifractal pattern obtained in real or computer experiments, one frequently employs the so-called {\it box counting} procedure which can be briefly described as follows.  Subdivide the sample into $M_l=(L/l)^d$ non-overlapping hypercubic boxes $\Omega_k$ of linear dimension $l$.  Associate with each box the mean height $H_k(l)=(l/a)^{-d}\sum_{i\in \Omega_k} h_i$, and define the scale-dependent partition functions
\be\label{scaledep}
Z_q(l,L)=\frac{1}{M_l}\sum_{k=1}^{M_l} \left[H_k(l)\right]^q
\ee
Note that for $l=a$ obviously $M_a=M=(L/a)^d$ and $Z_q(a,L)$ coincides with the partition function $Z_q$  featuring in the thermodynamic formalism. One may however observe that in the range $a\ll l \ll L$ the scale-dependent partition functions are sensitive to the spatial correlations in the heights at different lattice points. In particular, a simple consideration shows that when the heights are powerlaw-correlated in space as is actually the case for many systems of interest, see \cite{DL} and also below,  the scaling behaviour of $Z_q(l,L)$  depends non-trivially on both $l/a$ and $L/a$. At the same time the behaviour of the combination $I_q(l,L)=Z_q(l,L)/\left[Z_1(l,L)\right]^q$ turns out to be a function only on the scaling ratio $L/l$ and is given by
 \be\label{IPR}
 I_q(l,L)=\frac{Z_q(l,L)}{\left[Z_1(l,L)\right]^q} \sim \left(\frac{L}{l}\right)^{-\tau_q}, \quad\mbox{where} \quad \tau_q=d(q\zeta_1-\zeta_q)
\ee
which allows to get reliable numerical values of the scaling exponents $\tau_q$ by varying the ratio $L/l$ over a big range.
Further noticing that for $q=1$ the $l-$dependence of the partition function disappears due to linearity: $Z_1(l,L)=(l/a)^{-d}(L/l)^{-d}\sum_{1}^M h_i\sim (L/a)^{\zeta_1-d}$ we also can reliably extract $\zeta_1$ from the same data, hence
relate the set of exponents $\tau_q$ to $\zeta_q$ for $q\ne 1$.

The quantities $I_q=I_q(a,L)$ have interpretation of the {\it inverse participation ratio's} (IPR's) and are very popular in the theory of
the Anderson localization \cite{EM} and related studies. Passing from the partition functions $Z_q$ of the thermodynamic formalism to the IPR's is equivalent to focusing on the properties of the normalized {\it probability measure} $0<p_i=h_i/Z_1<1, \, \sum_i p_i=1$ rather than on the original height pattern $h_i$ itself. In fact in such a setting it is more natural to introduce the scaling of those weights in the form $p_i\sim M^{-\alpha_i}, \, \alpha_i\ge 0$ and consider the corresponding singularity spectrum $f(\alpha)$ related directly to the Legendre transform
of the exponents $\tau_q$. Working with the exponents $\tau_q$ has some advantages, as one can show they must be monotonically increasing convex function of $q$: $\frac{d\tau_q}{dq} >0, \, \frac{d^2\tau_q}{dq^2} \le 0$. In many situations, as e.g.  the diffusion-limited aggregation\cite{DLA} or indeed the Anderson localization the multifractal probability measures arise very naturally. In other contexts, e.g. in turbulence or in financial data analysis, the normalization condition seems superfluous. In the main part of the present paper we are mainly interested in the pattern of heights and are therefore concentrating on partition functions. We will discuss the normalized multifractal probability measures and associated IPR's briefly in the end.

The major features of the picture outlined above is of general validity for a given single multifractal pattern of any
nature, not necessarily random. In recent years considerable efforts were directed towards understanding
disorder-generated multifractality, see e.g. \cite{EM,FRL,ME,BogomGiraud}
 for a comprehensive discussion in the context of
Anderson localisation transitions and various associated random matrix models,
\cite{Dupl2000}, \cite{RBGW} in the context of harmonic measure generated by conformally invariant two-dimensional random curves
and \cite{Dirac,MG,F09} for examples
related to Statistical Mechanics in disordered media. We just briefly mention here that one of the specific
features of multifractality in the presence of disorder is a
possibility of existence of two sets of exponents,
$\tau_q$ versus $\tilde{\tau}_q$, governing the scaling behaviour of the
typical IPR denoted $I_q^{(t)}\sim
M^{-\tau_q}$ versus disorder averaged ("annealed") IPR, $\overline{I_q}\sim
M^{-\tilde{\tau}_q}$. Here and henceforth the overline stands for the averaging over
different realisations of the disorder. Namely, it was found that for large enough $q>q_c$ the two exponents will have in general different values: $\tau_q\ne \tilde{\tau}_q$.
The possibility of "annealed" average to produce results different from typical is related
to a possibility of disorder-averaged moments to be dominated by exponentially rare configurations.
As a result, the part of the "annealed" multifractality spectrum recovered via the Legendre transform from $\tilde{\tau}_q$ for
 $q>q_c$  will be negative \cite{negf},\cite{EM}: $\tilde{f}(x)<0$ for $x<x_{-}$, and similarly for  $x>x_{+}$.
 Further detail can be found in the cited papers and in the lectures \cite{F10}.

 Another important aspect of random multifractals revealed originally by Mirlin and Evers \cite{ME} in the context of the Anderson localization transition is the fact that IPR's $I_q$ for disorder-induced multifractal probability measures are generically power-law distributed: ${\cal P}(I_q/I_q^{(t)})\sim (I_q/I_q^{(t)})^{-1-\omega_q}$ \cite{ME,EM}.
   Such behaviour suggests that the actual values of the counting functions $N_{>}(x), N_{<}(x)$ should also show substantial sample-to-sample fluctuations, even in the range  $x_{-}<x<x_{+}$ where the singularity spectrum $f(x)$ is self-averaging and the same multifractal scalings $N_{>}(x)\sim M^{f(x)}$ is  to be observed in every realization of the pattern. Though the presence of such fluctuations was already mentioned in \cite{Dirac}, a detailed quantitative analysis seems to be not available yet. The main goal of our paper is to achieve a better understanding of statistics of the counting functions $N_{>}(x), N_{<}(x)$ by performing a detailed analytical as well as numerical study of arguably the simplest, yet important class of multifractal disordered patterns - those generated by one-dimensional Gaussian processes with logarithmic correlations, the so-called  $1/f$ noises.

   The structure of the paper is as follows. In the next section we will introduce the $1/f$ noise signals and discuss their properties already known
  from the previous works. Then we will use that knowledge to show that the probability
  density of the counting function $N_{>}(x)$ for such a model is characterized by a limiting scaling law with a powerlaw forwards tail, with the power governing the decay changing with the level $x$. We will then demonstrate that such powerlaw decay has nontrivial implications for the position of the maxima (or, with due modifications, minima) of such processes, and derive the expression for the threshold of extreme values. Finally, using $1/f$ noises as guiding example we will attempt to reinterpret the results of the theory developed in \cite{ME} to get a rather general prediction for the position of extreme value threshold for a broad class of disorder-generated multifractal patterns whose intensity is
  characterized by power-law correlations. We conclude with briefly discussing a few open questions.

  \section{$1/f$ noise: mathematical model and previous results}

An ideal $1/f$ (or "pink") noise is a random signal such that spectral power (defined via the Fourier transform of the autocorrelation function of the signal) associated with a given Fourier harmonic is inversely proportional to the frequency $\omega=2\pi f$.
Signals of similar sort are known for about eighty years and believed to be ubiquitous in Nature, see \cite{1frev} for a discussion and further references. Rather accurate $1/f$ dependences may extend for several decades in frequency in some instances, es. e.g. in voltage fluctuations in thin-film resistors \cite{1fres} or
resistance fluctuations in single-layer graphene films \cite{graphene},
 in non-equilibrium phase transitions \cite{1fnoneq}, and in spontaneous brain activity \cite{1fbrain}. Still, the  physical mechanisms behind such a behaviour are not yet fully known, and are a matter of active research and debate. It was noticed quite long ago (see e.g. \cite{HB}) that a generic feature of all such signals is that the two-point correlations (covariances) depend {\it logarithmically} on the time separation. During the last decade it became clear that random functions of such type appear in many interesting problems of quite different nature, featuring in
physics of disordered systems \cite{Dirac},\cite{CLD},\cite{FB},\cite{FLDR2009},\cite{FLDR2010}, quantum chaos \cite{1fchaos},
mathematical finance \cite{BKM,saakian}, turbulence \cite{turb1,MBP,turb2} and related models \cite{BM,Ostr,OstrRev}, as well as in mathematical studies of random conformal curves \cite{welding}, Gaussian Free Field  \cite{BDG,D} and related models inspired by applications is statistical mechanics \cite{AZ} and quantum gravity \cite{DRSV}, and the most recently in the value distribution of the characteristic polynomials of random matrices and  the Riemann zeta-function along the critical axis \cite{FHK}.
Let us note that a simple argument outlined in \cite{F10} and repeated in section 5 of the present paper shows that
 by taking the logarithm of any spatially homogeneous powerlaw-correlated multifractal random field we necessarily obtain a field logarithmically correlated in space. A somewhat similar in spirit suggestion to refocus the attention from the multifractal ("intermittent") signals and fields to their logarithms was also put forward in \cite{MBP}. All this makes logarithmically-correlated Gaussian processes an ideal laboratory for studying disorder-induced multifractality, though
 investigating the effects of non-Gaussianity remains a challenging outstanding issue.

Despite the fact of being of intrinsic interest and fundamental importance,
a coherent and comprehensive  description of statistical characteristics of ideal $1/f$ noises
  seems not to be yet available, and relatively few properties are firmly established even
  for the simplest case of a Gaussian $1/f$ noise. Among the works which deserve mentioning in such a context is the paper \cite{1fpower} which provided an explicit distribution of the "width" (or "roughness") for such a signal, as well as the work \cite{1ftrunc} describing a curious
  property of spectral invariance with respect to amplitude truncation. In recent papers \cite{FB},\cite{FLDR2009} and \cite{AZ}  the statistics of the {\it extreme} (minimal or maximal) values of various versions of the ideal $1/f$ signals was thoroughly addressed. From that angle the subject of the present paper is to provide a fairly detailed picture of statistics of the number of points in such signals which lie  above a given threshold set at some {\it high}  value. The latter can be rather naturally defined as being at finite ratio to the typical value of the absolute maximum.

 In this paper we are going to consider only Gaussian ideal $1/f$ noises, $2\pi$-periodic version of which is naturally defined via a random Fourier series of the form
 \be\label{1fperiodic} V(t)=\sum_{n=1}^{\infty}
\frac{1}{\sqrt{n}} \left[v_n e^{i n t}+v^*_n e^{-i n t}\right]
 \ee
 where  $v_n$ is a set of i.i.d. complex Gaussian variables with zero mean and the variance  $\overline{ |v_n|^2}=1$,
 with the asterisk standing for the complex conjugation, and the bar for the statistical averaging. It implies the following covariance structure
 of the random signal
 \be\label{1fperiodicvar}
\overline{V(t_1)V(t_2)}=-2\ln|2\sin{\frac{t_1-t_2}{2}}|, \quad t_1\ne t_2\in[0,2\pi)
\ee
Mathematically such series represents the periodic version of the fractional Brownian motion  with the Hurst index $H=0$. The corresponding definition is formal, as the series in (\ref{1fperiodic}) does not converge pointwise, the fact reflected, in particular, in the logarithmic divergence of the covariance in (\ref{1fperiodicvar})\footnote{One can also define other, in general non-periodic versions of similar log-correlated random processes on finite intervals using different basis of orthogonal functions, or even exploit the appropriate random Fourier integral to define the process on the half-line $0<t<\infty$. The corresponding models arise very naturally in the context of Random Matrix Theory and will be discussed in a separate publication \cite{FyoKhorSim}.}. Although it is possible to provide several {\it bona fide} mathematically correct definitions of the ideal $1/f$ noise as a random generalized function (based, for example, on sampling $2d$ Gaussian free field along the specified curves , e.g. the unit circle for the periodic noise, see \cite{welding}, or the constructions proposed in  \cite{BM} or \cite{AZ}), for all practical purposes the $1/f$ noises should be understood after a proper regularization. In what follows we will use explicitly the regularization proposed by Fyodorov and Bouchaud \cite{FB},
though we expect the main results must hold, {\it mutatis mutandis}, for any other regularization.

 In the model proposed in \cite{FB} one subdivides the interval $t\in (0,2\pi]$ by finite number $M$ of observation points $t_k=\frac{2\pi}{M}k$ where $k=1,\ldots,M<\infty$, and replaces the function $V(t), \, t\in[0,2\pi)$ with a sequence of $M$ random mean-zero Gaussian variables $V_k$ correlated according to the $M\times M$  covariance matrix $C_{km}=\overline{V_k V_m}$ such that
 the off-diagonal entries are given by
\be \label{logcovardisc}
C_{k\ne m}=-2\ln{|2\sin{\frac{\pi}{M}(k-m)}|}\,.
\ee
To have a well-defined set of the Gaussian-distributed random variables one has to ensure the positive definiteness of the covariance matrix
by choosing the appropriate diagonal entries $C_{kk}$. A simple calculation \cite{FB} shows that as long as we choose
 \be \label{logvardisc}
 C_{kk}=\overline{V_k^2}>2\ln{M}, \quad \forall k=1,\ldots, M \,.
\ee
 the model is well defined, and we will actually take the minimal possible value: $ C_{kk}=2\ln{M}+\epsilon,\, \forall k$ with a small positive $\epsilon\ll 1$. We expect that the statistical properties of the sequence $V_k$ generated in this way reflect correctly the universal features of the $1/f$ noise. An example of the signal generated for $M=4096$ according to the prescription above  via the Fast Fourier Transform (FFT) method as explained in detail in \cite{FLDR2009} is given in the figure.

 \begin{figure}[h!]
 \begin{center}
 \includegraphics[width=10cm]{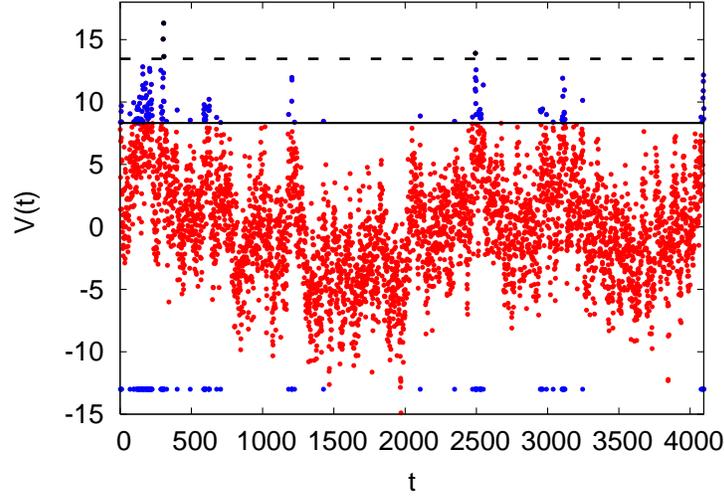}
\end{center}
\caption{A sample of the regularized periodic $1/f$ noise for $M=4096$ observation points. The upper broken line marks the typical value of the maximum $V_m=2\ln{M}-\frac{3}{2}\ln{\ln{M}}$ and black dots mark a few exceedances of that level. The lower solid line is at the  level $\frac{1}{\sqrt{2}}V_m$ and the blue dots at the bottom mark points $i$ supporting $V_i>\frac{1}{\sqrt{2}}V_m$. The set looks like a fractal.}
\end{figure}

Using the model (\ref{logcovardisc})-(\ref{logvardisc}) the authors of \cite{FB} defined the  associated random energy model via the partition function $Z(\beta)=\sum_{i=1}^Me^{-\beta V_i}$, with the temperature $T=\beta^{-1}\ge 0$ and succeeded in determining the distribution of $Z(\beta)$ in the range $\beta<1$.
To reinterpret those findings in the context of multifractality we introduce the height variables $h_i=e^{V_i}>0$ and rename
 $\beta\to -q$ , converting $Z(\beta)$ of the random energy model to $Z_q$ of the "thermodynamic formalism", eq.(\ref{1}).
 Note that due to the statistical equivalence of $V_i$ and $-V_i$ in the model all results may depend only on $|q|$.
Then the findings of \cite{FB} can be summarized as follows. The probability density of the random variable $Z_{|q|<1}$
consists of two pieces, the {\em body} and the {\em far tail}.  The  body of the distribution has a pronounced maximum at $Z \sim Z_e(q)=M^{1+q^2}/\Gamma(1-q^2)\ll M^2$, and a powerlaw decay when $Z_e\ll  Z\ll M^2$. Introducing $z=Z_q/Z_e(q)$ the probability
density of such a variable is given explicitly by
\be \label{partdissbody}
 {\cal P}_q(z)=\frac{1}{q^2}\left(\frac{1}{z}\right)^{1+\frac{1}{q^2}}\,
e^{-\left(\frac{1}{z}\right)^{\frac{1}{q^2}}},\quad  z\ll M^{1-q^2}, \, \quad |q|<1
\ee
 For $z\gg  M^{1-q^2}$ the above expression is replaced by a lognormal tail \cite{FB}.
 Note that the probability density (\ref{partdissbody}) is characterized by the
 moments
\be \label{moments}
\overline{z^s}_{M\gg 1}\approx  \, \Gamma(1-s q^2), \quad  \mbox{Re}\,s<q^{-2}
\ee
where $\Gamma(z)$ is the Euler Gamma-function.
It is worth noting that although the particular form of the density (\ref{partdissbody}) is specific for the chosen model
of $1/f$ noise, the power-law forward tail  ${\cal P}_q(Z)\propto Z^{-1-\frac{1}{q^2}}$  is expected to be universal \cite{FLDR2009}
and so the divergence of moments of the partition function for $\mbox{Re}\,s>q^{-2}$. A closely related fact which can be also traced back to the existence of the universal forward tail is that the typical  partition function scale $Z_e(q)$  in all $1/f$ models is expected to behave for $q\to 1$ as $Z_e(q)/M^{1+q^2}\sim (1-q)\to 0$. This property will have important consequences at the level of the counting function.

\begin{figure}[h!]
 \begin{center}\label{figz}
 \includegraphics[width=12cm]{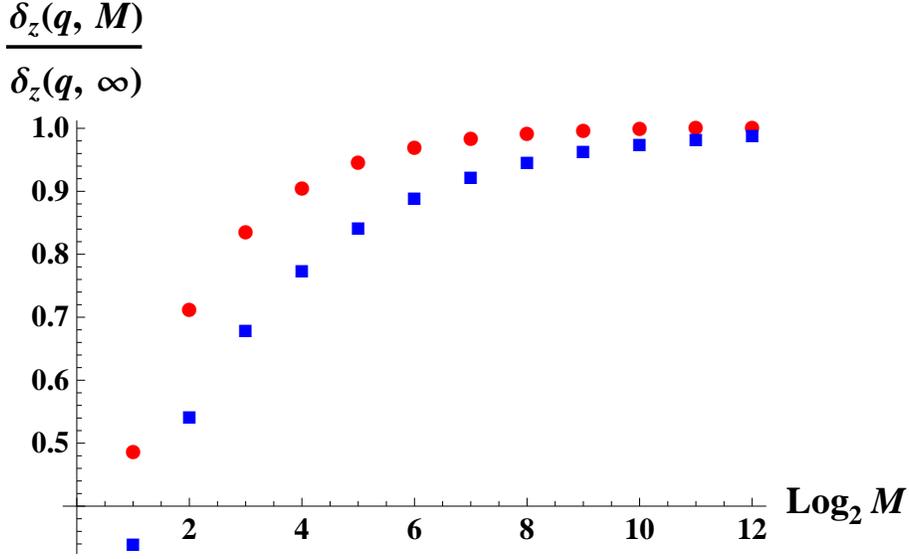}
\end{center}
\caption{ Convergence of the relative variance to the asymptotic value.
Red dots correspond to $q=1/3$ and blue dots correspond to $q=1/2$. The asymptotic behavior is reached for $M\sim 2^{11}$. When $q$ increases the convergence slows down.}
\end{figure}

To get some understanding of how the above asymptotic results agree with the direct numerical simulations
of the model (\ref{logcovardisc})-(\ref{logvardisc}) for large, but finite $M$ it is useful to provide the exact finite-$M$ expression for the second moment of the partition function, see \cite{FB}:
\be\label{exacvarz}
\fl \overline{Z_q^2}=M^{1+4q^2}+M^{2(1+q^2)}\,S_2^{(M)}(q^2), \quad S_2^{(M)}(q^2)=\frac{1}{M}\sum_{l=1}^{M-1}\left[4\sin^2{\left(\frac{\pi}{M}l\right)}\right]^{-q^2}
\ee
Second term here is dominant in the large-$M$ limit and gives precisely the result (\ref{moments}) as
asymptotically we can replace the sum by the integral convergent for $q^2<1/2$ and get
\be\label{beta-integral}
\lim_{M\to \infty} S_2^{(M)}(q^2)=\frac{2}{\pi}\int_0^{\pi/2}\left[4\sin^2{\theta}\right]^{-q^2}\,d\theta=
\frac{\Gamma(1-2q^2)}{\Gamma^2(1-q^2)}
\ee
The main correction to this asymptotic result is given by the first term in (\ref{exacvarz}), whose relative contribution is small
as $M^{1-2q^2}$ for $q^2<1/2$. In figure \ref{figz} we show the exact evaluation  for the relative variance $\delta_z(q,M)=\frac{\overline{z^2}}{\left(\overline{z}\right)^2}-1$ with  $z=Z_q/Z_e(q)$ given explicitly by
\be\label{exacdeltaz}
 \delta_z(q,M)= S_2^{(M)}(q^2) +M^{2 q^2-1} -1
\ee
We see that  $ \delta_z(q,M)$  clearly approaches the asymptotic value $\delta_z(q,\infty)=\frac{\Gamma(1-2q^2)}{\left[\Gamma(1-q^2)\right]^2}-1$ for $M\sim 2^{11} $.

\section{ Thermodynamic formalism for the counting function of the $1/f$ noise sequence and the threshold of extreme values}

 The statistics of $Z_q$ for the model under consideration suggest that it reflects the corresponding strong sample-to-sample fluctuations in the counting function of pattern of heights. Our goal is to quantify statistics of those fluctuations by considering the total number $N_>(x)$, which in
the present context will be denoted as  ${\cal N}_M(x)$, of the  {\it $x-$high} points in the (regularized) $1/f$ sequence $V_1,\ldots, V_M$ .
Those points are defined as such that $V_i>x\ln{M}$ which is equivalent to $h_i=e^{V_i}>M^{x}$.

\begin{figure}[h!]
\label{figmeanvstyp}
 \includegraphics[width=10cm]{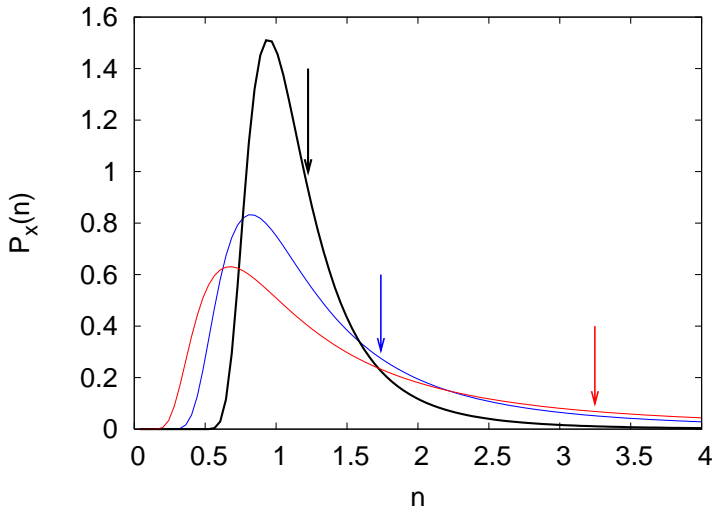}
\caption{   From top to bottom the probability density Eq.(\ref{dissnumb}) for $x=1$, $x=1.4$, $x=1.7$. Arrows indicates the position of
the mean, the typical value corresponds always to $1$ in the chosen scaling. }
\end{figure}

 We then relate the number ${\cal N}_M(x)$ to the partition function $Z_q$  by the thermodynamic formalism:
\be\label{defcount}
 {\cal N}_M(x)=\ln{M} \int_x^{\infty}\,\tilde{\rho}_M(y)\,dy,\,\quad Z_q=\ln{M} \int_{-\infty}^{\infty}M^{q y}\tilde{\rho}_M(y)\,dy\,
\ee
where now the density $\tilde{\rho}_M(y)=\frac{\rho_M(y)}{\ln{M}}=\sum_{k=1}^M\delta(V_k-y\ln{M})$ is anticipated to be given in the large-$M$ limit by the multifractal ansatz of the particular "improved" form:
\be\label{multiansatzdenhigh}
 \tilde{\rho}_M(y)\approx c_M(y) \frac{M^{1-y^2/4}}{\sqrt{\ln{M}}}, \quad c_M(y)=\frac{n_M(y)}{2\sqrt{\pi}\Gamma(1-y^2/4)}, \quad |y|<2\,.
\ee
Here $n_M(y)$ is assumed to be a random coefficient of order of unity
which strongly fluctuates from one realization of the sequence $V_i$ to the other
in such a way that its probability density is given by  the formula (\ref{partdissbody}) with $q$ value chosen to be $q=y/2$.
Indeed, substituting the density (\ref{multiansatzdenhigh}) to $Z_q$ in (\ref{defcount}) and performing the integrals in the limit $\ln{M}\gg 1$ by the Laplace method we arrive at the asymptotic behaviour $Z_q\approx n_M(2q) \, Z_e(q), \,\, q<1$, with the value $Z_e(q)=\frac{M^{1+q^2}}{\Gamma(1-q^2)}$ precisely as we have found from the exact solution (\ref{partdissbody}).  On the other hand, substituting the same ansatz to the counting function in (\ref{defcount}) yields by the same method ${\cal N}_{M\gg 1}(x)\approx  n_M(x)\, {\cal N}_t(x)\,$ where the typical value ${\cal N}_{t}(x)$ is given by
\be\label{charscale}
{\cal N}_{t}(x)= \frac{M^{1-x^2/4}}{x\sqrt{\pi\ln{M}}}\frac{1}{\Gamma(1-x^2/4)}, \quad 0<x<2\,.
\ee
Thus, our main conclusion is that two random variables $n={\cal N}_M(x)/{\cal N}_{t}(x)$
and  $z=Z_{q/2}/Z_{e}(q/2)$  must be distributed in the large$-M$ limit according to the same probability law, which after invoking (\ref{partdissbody}) yields the asymptotic probability density for the scaled counting function in the form
\be \label{dissnumb}
 {\cal P}_x(n)=\frac{4}{x^2}\frac{1}{n^{1+\frac{4}{x^2}}}\,
e^{-\left(\frac{1}{n}\right)^{\frac{4}{x^2}}},\quad 0<x<2\,.
\ee
The shape of the distribution for a few values of $x$ is presented in Fig. \ref{figmeanvstyp}.

The following qualification is needed here. For large but finite $M$ such form of the density stops to hold true for extremely large $n\to \infty$ as in any realization obviously ${\cal N}_M(x)<M$ at the very least. Therefore there must exist an upper cut-off value ${\cal N}_c(x)$ such that for $n>n_c={\cal N}_c(x)/{\cal N}_t(x)$ the scaling form of the probability density (\ref{dissnumb}) loses its validity. The cutoff $n_c$ should diverge
as long as $M\to \infty$. Similarly, another restriction on the validity of (\ref{dissnumb}) should exist in the region
of  extremely small $n\to 0$ due to implicit condition ${\cal N}_M(x)\gg 1$. Precise value of the cutoffs can not be extracted in the framework of the thermodynamic formalism, and its determination remains an open issue.

The important scale ${\cal N}_{t}(x)$  defined explicitly in  (\ref{charscale})
 describes {\it typical} values of the counting function ${\cal N}_M(x)$ for a given observation level $x$. In particular, it can be used to define
  one of the objects of central interest in the present paper, the
  {\it threshold of extreme values}. The latter stands for such a level above which typically we can find for $\ln{M}\gg 1$ only a few, i.e. of the order of one points of our random sequence.
 The scaling behaviour ${\cal N}_{t}(x)\sim M^{f(x)}, \, f(x)=1-x^2/4$
 is the hallmark of the multifractality. A very similar parabolic singularity spectrum characterizes the high value pattern of the two-dimensional Gaussian free field as revealed in \cite{Dirac} and proved in a mathematically rigorous way in
\cite{D}. The above result for $f(x)$ is the simple one-dimensional analogue of that fact, see e.g. \cite{AZ} where it is rigorously shown that $\lim_{M\to \infty}\frac{\ln{{\cal N}_M(x)}}{\ln{M}}=1-x^2/4$ in our notations.  Note that as
 the singularity spectrum $f(x)$ vanishes at $x=2$ the typical position of the absolute maximum of the random sequence of $V_i$'s is given by $V_{max}=2\ln{M}$ at the leading order. The corresponding subleading term was conjectured in the work by Carpentier and Le Doussal \cite{CLD} to be $V_m=2\ln{M}-c\ln{\ln{M}}+O(1)$, with $c=3/2$. That conclusion was based upon an analysis of the travelling wave-type equation \cite{DS} appearing in the course of one-loop renormalization group calculation, and the value $c=3/2$ was conjectured to be universally shared by all systems with logarithmic correlation. Such a result
  is markedly different from $c=1/2$ typical for short-ranged correlated random signals, so the value of $c$ may be used as a sensitive indicator of the universality class. Indeed, in a recent numerical studies of the behaviour of the logarithm of the modulus of the Riemann zeta-function along  the critical line \cite{FHK} the value $3/2$ was used to confirm the consistency of describing that function as a representative of logarithmically correlated
  processes. Despite its importance, no transparent qualitative argument explaining $c=3/2$ vs. $c=1/2$  values was ever provided, to the best of our knowledge, though for the case of the 2D Gaussian free field the value $3/2$ was very recently rigorously proved by Bramson and Zeitouni \cite{BZ} by exploiting elaborate probabilistic arguments. Below we suggest a very general and transparent argument showing that the change from $c=1/2$ to $c=3/2$ is a direct consequence of the strong fluctuations in the counting function reflected in the power-law decay of the probability density (\ref{dissnumb}). That observation not only allows one to explain the origin of $c=3/2$ for the Gaussian case, but has also a predictive power
  in a more general situation as will be demonstrated in the section 5.

 To begin with presenting the essence of our argument we first observe that (\ref{dissnumb}) implies that $\overline{n}=\Gamma\left(1-\frac{x^2}{4}\right),\, 0<x<2$ so that the mean value of the counting function is given asymptotically by
\be\label{meanasymp}
\overline{{\cal N}_M(x)}\approx
  \frac{M^{1-x^2/4}}{x\sqrt{\pi\ln{M}}},
\ee
 We shall see in the next section that the above expression is asymptotically exact for any real $x>0$ without restriction to $0<x<2$.
 Notice however that the mean value (\ref{meanasymp}) and the characteristic scale $N_{t}(x)$ in (\ref{charscale}) differ from each other by the factor $\frac{1}{\Gamma(1-x^2/4)}$ tending to zero as $x\to 2$.  Such a difference, origin of which can be again traced back to
  the specific power-law tail of the probability density (\ref{dissnumb}), see Fig. \ref{figmeanvstyp}, is one of the hallmarks of the random signals with logarithmic correlations. Indeed, consider for comparison the case of uncorrelated i.i.d. Gaussian sequence sharing the same variance $\overline{V_i^2}=2\ln{M}$ with the logarithmically correlated noise (by historical reasons it is natural to refer to such model as the Random Energy Model, or REM \cite{REM}).
   A straightforward calculation shows that we still would have precisely the same mean value (\ref{meanasymp}) of the counting function as in the logarithmically-correlated case, but unlike the latter it will be simultaneously the typical value of that random variable as no
   powerlaw tail is present in that case (see Fig. \ref{figpn} and discussion in the next section).

Such a difference between the two cases has important implications for the location of the threshold $x=x_m$ which corresponds to the region of extreme value statistics of multifractal heights.
Indeed, by approximating the singularity spectrum $f(x)$ close to its right zero $x_{+}=2$ as $f(x)\approx (2-x)$, and similarly writing
 $\Gamma\left(1-\frac{x^2}{4}\right)\propto (2-x)^{-1}$ we observe that the condition ${\cal N}_t(x_m)\sim 1$ is equivalent to the equation:
\be\label{thresholdcond}
(2-x_m)\ln{M}-\frac{1}{2}\ln{\ln{M}}+\ln{(2-x_m)}=0
 \ee
 solving which for $\ln{M}\gg 1$ to the first non-trivial order gives precisely $x_m=2-c\,\frac{\ln{\ln{M}}}{\ln{M}}$ with $c=3/2$.
Had we replaced in the above condition the typical value ${\cal N}_t(x_m)$  with the mean  $\overline{{\cal N}_M(x_m)}$ we would arrive to the
same expression for $x_m$ but with the value $c=1/2$ replacing $c=3/2$. This perfectly agrees with such $x_m$ being the extreme value threshold for short-ranged correlated random sequences. The suggested explanation of the transmutation of the coefficient $c$ based on the "typical versus mean" argument seems to us very transparent and supports the conjectured universality of the result. Indeed, the thermodynamic formalism combined with the results of  \cite{FLDR2009} suggests that the power-law forward tail  ${\cal P}_x({\cal N})\propto \frac{1}{{\cal N}}\left(\frac{{\cal N}_e}{{\cal N}}\right)^{\frac{4}{x^2}}$ should be universal for one-dimensional Gaussian processes with logarithmic correlations. It also should show up, {\it mutatis mutandis},  in higher-dimensional versions of the model like e.g. the Gaussian Free Field on the lattice. Such a tail will ensure the difference between the typical and the mean of the counting function by a factor which vanishes linearly on approaching the extreme value threshold value  $x=x_m$. This factor then will lead to the value $c=3/2$ by the mechanism illustrated above for our particular explicit example.
The case of non-Gaussian signals with logarithmic correlations is relevant for general disorder-generated multifractals and is discussed in Section 5 of the paper.

Finally it is worth mentioning that for $x>2$ the mean value of the counting function (\ref{meanasymp}) is exponentially small.   This fact reflects the need to generate exponentially large number of samples to have for $\ln{M}\gg 1$ at least a single event with $V_i>2x\ln{M}$ when $x>2$. Indeed, such values of $V_i$ will not show up in a typical realization (cf. earlier discussion about "annealed" vs. "quenched" singularity spectra).

\section{Exact results versus asymptotics}

\begin{figure}[h!]
\label{figpn}
 \includegraphics[width=10cm]{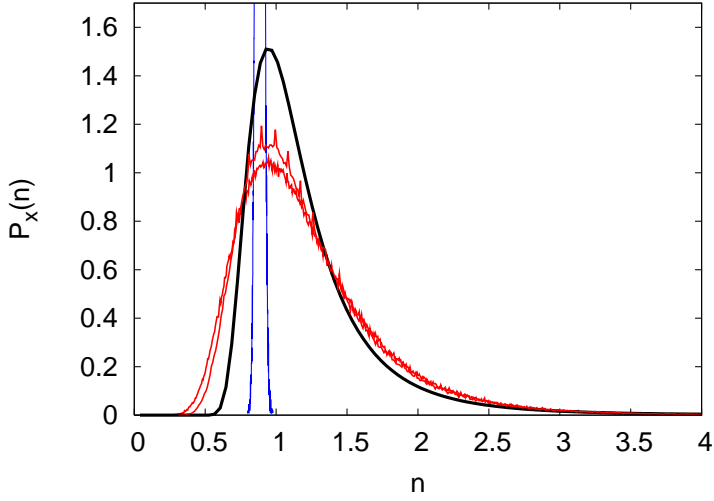}
\caption{  Probability density for the scaled counting function, $ {\cal P}_x(n)$ with  $x=1$. Black solid line corresponds to the analytical prediction Eq.\ref{dissnumb}.  Red lines correspond to numerical simulation of the regularized $1/f$ Gaussian signal generated via the Fast Fourier Transform (FFT) method
(lower red curve $M=2^{14}$, higher $M=2^{18}$, data collected from $10^5$ samples.). The blue line corresponds to the REM model: $M=2^{18}$ i.i.d. zero mean Gaussian variables with variance $2\ln{M}$; in that case the density for $n={\cal N}_M(x)/\overline{{\cal N}_M(x)}$ is clearly converging to the delta peak and does not show any power-law tails, see discussion in the text.}
\end{figure}

The above results for the counting function obtained in the framework of the thermodynamic formalism  are expected to be valid as long as $\ln{M}\gg 1$. To get a feeling of how big in practice should be $M$ to ensure the validity of
our asymptotic formulae it is natural again to try to perform the direct numerical simulations of the regularized version of the ideal $1/f$ noise.
We start with checking directly the distribution of the scaled counting function, Eq.(\ref{dissnumb}), for a particular value $x=1$.
The results are presented in Fig.\ref{figpn}. They show that although the main qualitative features of the distribution (in particular, the well-developed powerlaw tail) are clearly
in agreement with the theoretical predictions, the curve is still rather far from its predicted asymptotic shape for $M=2^{18}$, and the convergence is too slow to claim a quantitative agreement.

To get a better understanding of the mechanism of such disagreement at a quantitative level, and to check
 the results obtained in the framework of the thermodynamic formalism we choose to consider in much greater detail the first two moments of the counting function. The asymptotic formula (\ref{dissnumb})  yields for the mean of the counting function the expression (\ref{meanasymp}) and for the variance
\be\label{firsttwo}
 \frac{\overline{\left[{\cal N}_M(x)\right]^2}-\left[\overline{{\cal N}_M(x)}\right]^2}
 {\left[\overline{{\cal N}_M(x)}\right]^2} \approx \frac{\Gamma (1-x^2/2)}{\Gamma^2(1-x^2/4)}-1, \quad 0<x<\sqrt{2}
\ee
At the same time it is possible to derive a closed form exact finite $M$ expression for the first two moments of the counting function ${\cal N}_M(x)=\sum_{i=1}^M \theta(V_i-x\ln{M})$ without any recourse to the thermodynamic formalism. Here we have used the Heaviside
step function $\theta(u)=1$ for $u>0$ and $\theta(u)=0$ otherwise. The mean value can be immediately computed as
\be\label{meannumb}
\fl \overline{{\cal N}_M(x)}= \frac{M}{2\sqrt{\pi\ln{M}}}\int_{x\ln{M}}^{\infty}\exp{\left(-\frac{v^2}{4\ln{M}}\right)}\,dv=
 \frac{M}{2}\,\mbox{Erfc}\left(\frac{x}{2}\sqrt{\ln{M}}\right)
\ee
and is independent of the correlations. The problem of deriving a closed-form expression for the variance which is amenable to accurate numerical evaluation for very big $\ln{M}\gg 1$ is less trivial and may have an independent interest. Before presenting the results we find it most convenient to define the following object
\be\label{exvar1}
\Delta_M(x)=\overline{N_M(x)\left({N}_M(x)-1\right)}-\frac{M-1}{M}\left[\overline{{ N}_M(x)}\right]^2
\ee
in terms of which the relative variance is expressed as
 \be\label{relvar}
   \fl
  \frac{\overline{\left[{\cal N}_M(x)\right]^2}-\left[\overline{{\cal N}_M(x)}\right]^2}
 {\left[\overline{{\cal N}_M(x)}\right]^2}=\frac{1}{M}+\frac{1}{\overline{{\cal N}_M(x)}}+\delta_n(x;M),\quad \delta_n(x;M)=\frac{\Delta_M(x)}{\left(\overline{{\cal N}_M(x)}\right)^2}
  \ee
  $\Delta_M(x)$ is a convenient measure of correlation-induced fluctuations. Using the definition of the counting function we explicitly get:
\be
\fl \Delta_M(x)=\sum_{i\ne j} \left[\overline{\theta(V_i-x\ln{M}) \theta(V_j-x\ln{M})}- \overline{\theta(V_i-x\ln{M})}\cdot \overline{ \theta(V_j-x\ln{M})} \right]
\ee
  from which is evident that  $\Delta_M(x)$ vanishes for any i.i.d.  sequence. In the latter case
  the relative variance tends to zero in the large-$M$ limit assuming $\overline{{\cal N}_M(x)}\to \infty$ for $M\to \infty$.
   This simply means that in the i.i..d. case the variable ${\cal N}_M(x)/\overline{{\cal N}_M(x)}$ is self-averaging, i.e its limiting density approaches the Dirac delta-function, see Fig. \ref{figpn}.
    On the other hand, $\Delta_M(x)$ is formally different from zero for correlated variables.
   Nevertheless, using the general formalism exposed in the Appendix A one can satisfy oneself that for all stationary Gaussian sequences with correlations decaying fast enough to zero at big separations (e.g. as a power of the distance) the quantity $\delta_n(x;M)$ still tends to zero as $M\to \infty$. In contrast, we will see below that in the logarithmically correlated case $\delta_n(x;M)$ tends to a finite positive number for $M\to \infty$, and thus coincides with the leading behaviour of the variance of the counting function.

In the Appendix A we have derived the exact expression for $\Delta_M(x)$ for a general correlated Gaussian sequence.
 For the periodic $1/f$ noise sequence using (\ref{logcovardisc}) the result reads
\be\label{exvar3}
\fl
\Delta_M(x)=\frac{1}{\pi}M^{1-x^2/4}\sum_{n=1}^{M-1} \int_{h_n}^1\frac{d\tau}{1+\tau^2}e^{-\ln{M}\,x^2\tau^2/4}, \quad h_n=\sqrt{\frac{\ln{M}+\ln{|2\sin{\left(\frac{\pi n}{M}\right)|}}}{\ln{M}-\ln{|2\sin{\left(\frac{\pi n}{M}\right)|}}}}
\ee

In the figure we test its validity by comparing the results obtained for moderate value of $M$ by direct numerical simulations of the log-correlated sequences with the predictions of (\ref{exvar3}).

\begin{figure}[h!]
\label{fignumerical2}
\includegraphics[width=8cm,angle=-90]{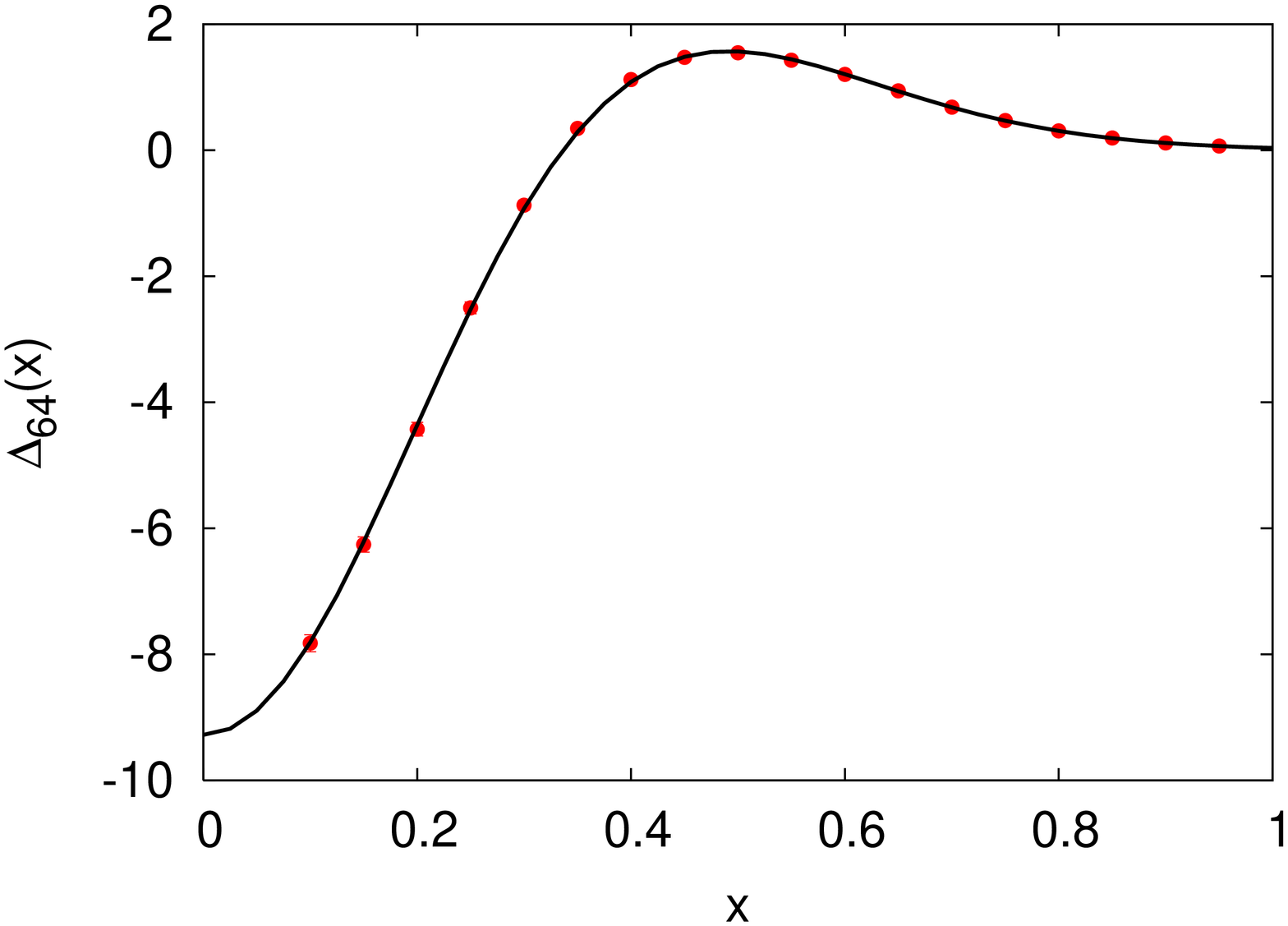} \includegraphics[width=8cm,angle=-90]{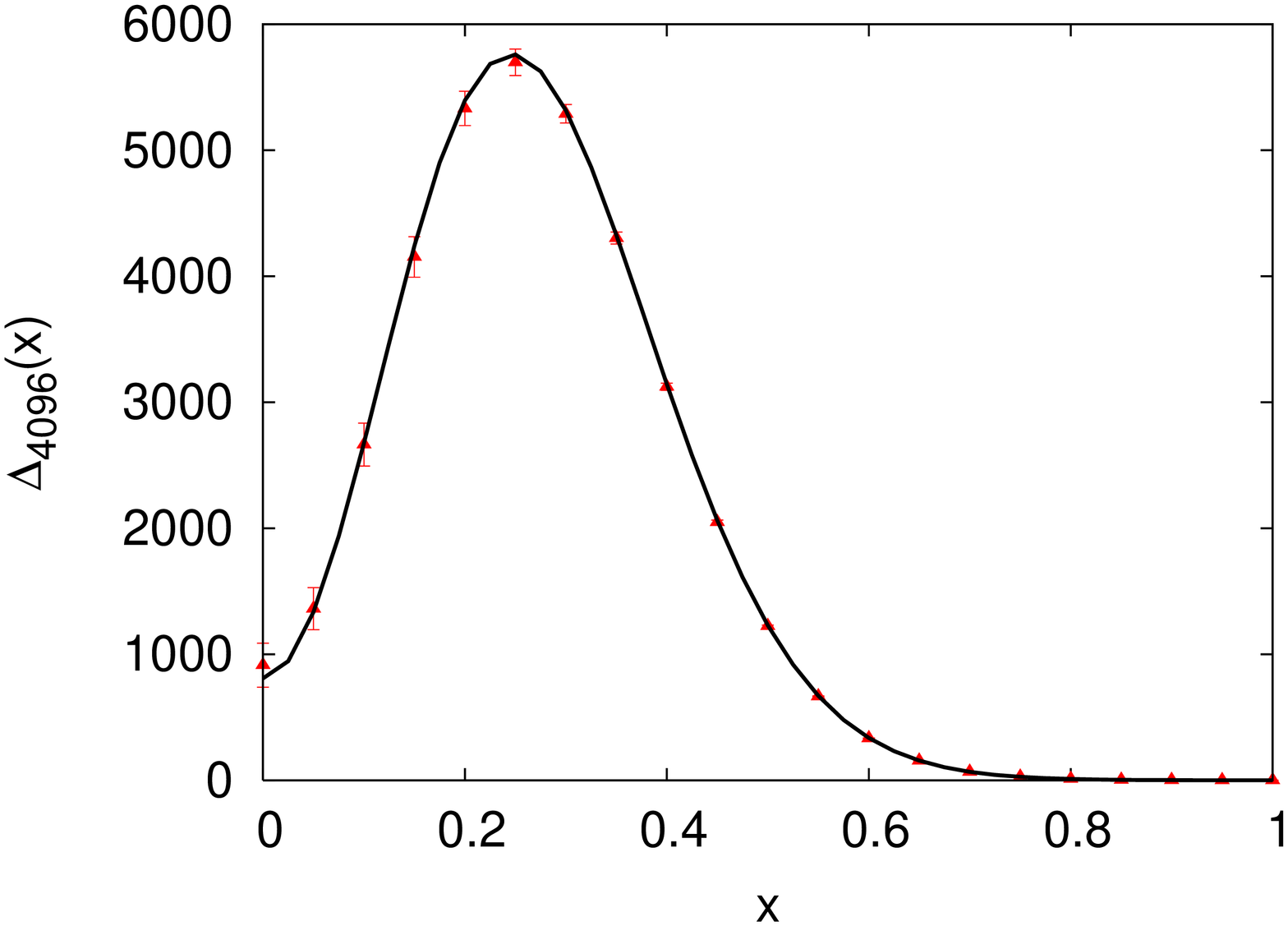}
\caption{Numerical simulations (red bars) compared to the theoretical prediction (\ref{exvar3}) (black line) for  $M=64$ (top) and  $M=4096$ (bottom). The $1/f$ signals are generated via the Fast Fourier Transform (FFT) method.}
\end{figure}

 The expressions above are suitable for developing a well-controlled approximation to the exact expression (\ref{exvar3}) in the large-$M$ limit assuming $\ln{M}\gg 1$. First of all, it is clear that in the large-$M$ limit we may replace the discrete sum in (\ref{exvar3}) by the integral
 treating $\frac{\pi n}{M}$ as a continuous  variable $\theta$. Using the symmetry $\theta\to \pi-\theta$ we then arrive to an approximation to the ratio $\delta_n(x;M)=\frac{\Delta_M(x)}{\left(\overline{{\cal N}_M(x)}\right)^2}$ given by
\be\label{approxdelta}
\delta_n(x;M)\approx \frac{8 M^{-x^2/4}}{\pi^2\mbox{Erfc}^2{\left(\frac{1}{2}x\sqrt{\ln{M}}\right)}}\int_{\frac{\pi}{M}}^{\frac{\pi}{2}}d\theta \int_{h_{\theta}}^1\frac{d\tau}{1+\tau^2}e^{-\ln{M}\,x^2\tau^2/4}\,\quad ,
\ee
where the lower limit of integration over $\tau$ is given by
\be\label{approxdelta1}
 h_{\theta}=\sqrt{\frac{\ln{M}+\ln{|2\sin{\theta}|}}
{\ln{M}-\ln{|2\sin{\theta}|}}}\,\quad.
\ee
At the next step we assume that the $\theta-$integral is dominated by the finite values $0<\theta<\pi/2$. This allows to replace
 the lower limit of integrations over $\theta$ by zero and also to expand $h_{\theta}\approx
 1+\ln{|2\sin{\theta}|}/\ln{M}$ for $\ln{M}\to \infty$. Changing then the integration variable $\tau=(1+u/\ln{M})^{1/2}$ and keeping only the leading order terms the $u-$integral is easily calculated and the emerging $\theta-$integral takes the form
 $\int_0^{\pi} \left(4\sin^2{\theta}\right)^{-x^2/4}\,d\theta$. The latter is convergent as long as $x<\sqrt{2}$ where it can be reduced to the Euler beta-function, see (\ref{beta-integral}). As a result, we arrive to the following expression:
\be\label{approxdeltalim}
\lim_{M\to \infty}\delta_n(x;M)=\frac{\Gamma\left(1-x^2/2\right)}{\Gamma^2\left(1-x^2/4\right)}-1, \quad 0<x<\sqrt{2}
\ee
which is fully equivalent to the variance result (\ref{firsttwo}) we have anticipated on the basis of the thermodynamic formalism.
For $x>\sqrt{2}$ the $\theta-$ integral diverges at the lower limit $\theta\to 0$  rendering our large-$M$ procedure invalid. This case will be separately treated in the end of the section.

 Now we are in a position to check numerically the range of applicability of the approximations derived. First we attempt to compare the results of exact numerical evaluation of the discrete sum in (\ref{exvar3}) to the integral (\ref{approxdelta}).
 Actually, the direct evaluation of the sum in Mathematica is affordable up to $M=50000$. To go up to the higher values of $M$ we use the identity
 \be\label{sumapprox}
 \sum_{n=1}^{M-1}\int_{h_n}^1 f(\tau)\,d\tau=\sum_{n=1}^{\frac{M}{2}-1}2R(n)\int_{h_n}^{h_n+1} f(\tau)\,d\tau
 \ee
 where $R(n)=n$ for $n<M/6$ and $R(n)=n-M/3+\frac{1}{2}$ for $n\ge M/6$. Since for large $\ln{M}$ the difference $|h_n-h_{n+1}|$ is very small, the
 integral in the above expression can be very accurately approximated by the trapezoidal rule
 \[
 2\int_{h_n}^{h_n+1} f(\tau)\,d\tau\approx (h_{n+1}-h_{n})\left(f(h_{n+1})-f(h_{n})\right)\,.
 \]
 This trick allows us to evaluate the sum in (\ref{exvar3}) numerically up to $M\sim 10^9$.

 \begin{figure}[h!]
 \label{fig4}
 \begin{center}
 \includegraphics[width=12cm]{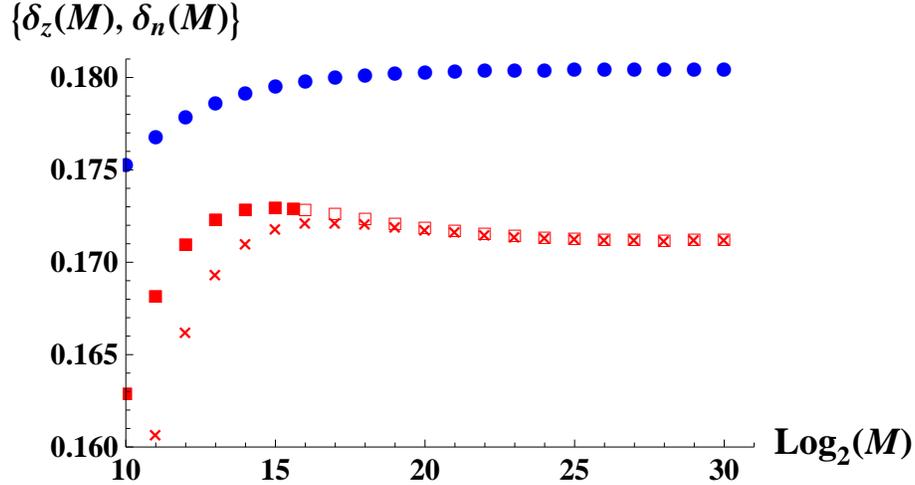}
\end{center}
\caption{Comparison of the convergence to the asymptotic values. The partition function variance (blue circles , $\delta_z(q=0.5,M)$) reaches the asymptotic value $\delta_z(q=0.5,\infty)=0.180341$ for $M\sim 2^{16}$. The relative variance  $\delta_n(x=1,M)$ for the counting function is presented as evaluated by three different methods: Filled Squares stand for the result of precise integration of the discrete sum (\ref{exvar3}) up to $M=50000$. Empty Squares stand for a numerical integration involving the identity (\ref{sumapprox}) and the trapezoidal rule. Diamonds describe the continuum approximation Eq.(\ref{approxdelta}). The asymptotic value predicted by (\ref{approxdeltalim}) for $M\to \infty$ is again $\delta_n(x=1;\infty)=0.180341$, and
 is nowhere close to the finite-$M$ results. The agreement even seems to worsen with growing $M$. See the next figure.}
\end{figure}

 The results are presented figure 5 and figure 6  for $x=1$. In figure 5 we compare the fast convergence of $\delta_z(q=0.5,M)$ against the very slow convergence of  $\delta_n(x=1;M)$. For $M\sim 2^{15} $
 $\delta_z(q=0.5,M)$  is already very close to the asymptotic value  whereas  $\delta_n(x=1,M)$ shows a curious non-monotonicity  and even for $M\sim 2^{30}$ we are still very far from
  the asymptotic value predicted by (\ref{approxdeltalim}) which is $\delta^{(n)}_{\infty}(x=1)=0.180341$.  We observe that for  $M$  larger than $ 2^{17}$ the continuum approximation of Eq.(\ref{approxdelta})  matches perfectly the  numerical integration  involving the identity (\ref{sumapprox}) and the trapezoidal rule. The latter  matches perfectly the exact discrete sum (\ref{exvar3}) even for moderate $M$.  As we now are confident in the accuracy of the
  continuum approximation formula (\ref{approxdelta}) given by a double integral with $\ln{M}$ entering as a simple parameter we can use it   to check what values of $\ln{M}$ actually ensure the validity of the asymptotic (\ref{approxdeltalim}). The results are presented in  fig. 6 and show that one needs astronomically big values of $M$ even to achieve a rather modest agreement with the asymptotic value.

\begin{figure}[h!]
\label{fig5}
 \begin{center}
 \includegraphics[width=12cm]{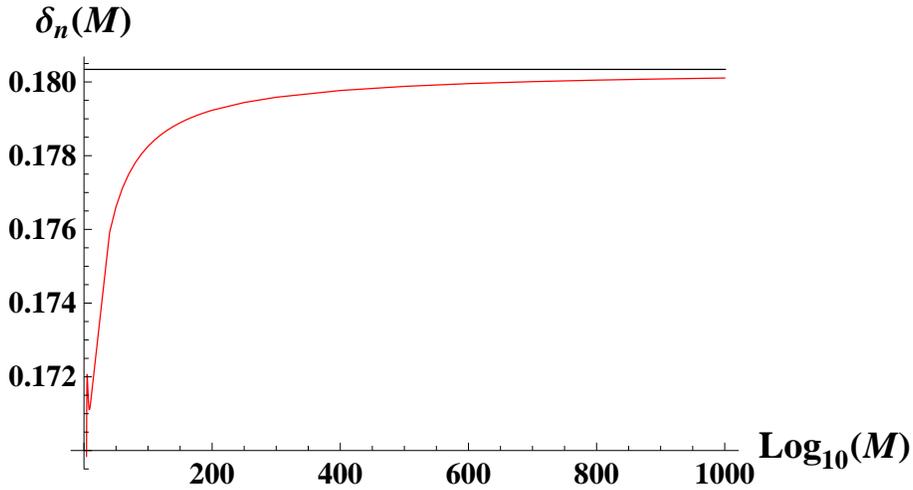}
\end{center}
\caption{$\delta_n(x=1,M)$ given by the continuum approximation formula (\ref{approxdelta}). The non-monotonic region seen in the previous figure is reflected in a tiny dip confined to a region close to the origin. The asymptotic value predicted by (\ref{approxdeltalim}) is $\delta_n(x=1;\infty)=0.180341$.}
\end{figure}

These facts explains our failure to confirm the infinite-$M$ asymptotic by direct simulation of the variance of the counting function predicted for the periodic $1/f$ noise in the framework of the thermodynamic formalism. In any realistic signal analysis of such variance
one must therefore rely upon the exact formula (\ref{exvar3}) or its analogues instead of the asymptotic value (\ref{approxdeltalim}).
The conclusion should be of significant practical importance, in particular in view of the growing interest in numerical investigations of
statistical properties of high values of the modulus of the Riemann zeta-function and of the characteristic polynomials of large random matrices.

 Having verified by the independent approach the asymptotic of the second moment of the counting function for $x<\sqrt{2}$ we can now
 apply a similar methods beyond that range. The formal divergence of the right-hand side in (\ref{approxdeltalim}) for $x\to \sqrt{2}$
 simply means that the second moment of the counting function is not proportional to the squared typical scale ${\cal N}_e$, but is parametrically larger. An accurate analysis of the second moment in the range $\sqrt{2}<x<2\sqrt{2}$ performed in the Appendix B shows that the leading asymptotic behaviour of $\delta_n(x;M)$ is given by
 \be\label{approx4}
\delta_n(x;M)\approx x^2\sqrt{\frac{\sqrt{2}\,\pi\ln{M}}{[2-x/\sqrt{2}]}}\,M^{\left(\frac{x}{\sqrt{2}}-1\right)^2}\,,
\quad \sqrt{2}<x<2\sqrt{2}
\ee
As is easy to see $\delta_n(x;M)\gg \left(\overline{N_M(x)}\right)^{-1}$ in the above domain, and is therefore the dominant term in the relative variance (\ref{relvar}).

\section{The position of threshold of extreme values in generic disorder-generated multifractal patterns}

 Results obtained so far in the paper suggest a natural question about statistics of high values and positions of extremes of more general
 power-law correlated  multifractal random field with a generic non-parabolic singularity spectrum. Most obvious examples include the variety of the Anderson transitions \cite{ME}, but in fact many more random critical systems should be of that sort, see e.g. \cite{DL,Dupl2000,RBGW,MG}.
 A straightforwards calculation outlined in \cite{F10} shows that behind each pattern of such type lurks a certain logarithmically
 correlated field, though in general of a non-Gaussian nature. Below we sketch that simple argument for the sake of completeness. Consider a $d-$dimensional sample of linear size $L$, and assume following \cite{DL} that the  multifractal patterns of intensities $p({\bf r})$ is {\it self-similar}
\be
\overline{p^q({\bf r_1})p^s({\bf r_2})}\propto L^{-y_{q,s}}|{\bf r}_1-{\bf r}_2|^{-z_{q,s}},\quad q,s\ge 0, \quad a\ll |{\bf r}_1-{\bf r}_2|\ll L
\ee
 and {\it spatially homogeneous}
\be
\overline{p^q({\bf r_1})}=\frac{1}{L^{d}}I_q,\quad \mbox{where}\quad I_q=\int_{|{\bf r}|<L} p^q({\bf r})\,d{\bf r}\propto L^{-\tau_q}
\ee
The consistency of the above two conditions for $|{\bf r}_1-{\bf r}_2|\sim  a$ and $|{\bf r}_1-{\bf r}_2|\sim  L$ implies:
\be
y_{q,s}=d+\tau_{q+s}, \quad z_{q,s}=d+\tau_{q}+\tau_{s}-\tau_{q+s}
\ee
If we now introduce the field $V({\bf r})=\ln{p({\bf r})}-\overline{\ln{p({\bf r})}}$ and combine the identity $\frac{d}{ds}p^s|_{s=0}=\ln{p}$ with the fact that $\tau_0=-d$ we straightforwardly arrive at the relation
\be
\overline{V({\bf r_1})V({\bf r_2})}=-g^2
\ln{\frac{|{\bf r}_1-{\bf r}_2|}{L}}, \quad g^2=\frac{\partial^2}{\partial s\partial q}\tau_{q+s}|_{s=q=0}
\ee
Thus we conclude that the logarithm of any multifractal intensity is a log-correlated random field. The above argument does not say
anything about higher cumulants of the field $V({\bf r})$, but it is easily checked that had those fields be always Gaussian the resulting singularity spectrum $f(\alpha)$ obtained from $\tau_q$ via the Legendre transform would be invariably parabolic. Therefore, any non-parabolicity of the singularity spectra necessarily implies non-Gaussian nature of the underlying log-correlated fields.
 Nevertheless, combining our previous insights with properties of disorder-generated multifractal patterns revealed in \cite{ME} suggests the way in which our results on Gaussian $1/f$ noise can be generalized to  statistics of high values and positions of extremes of more general non-Gaussian logarithmically correlated random processes and fields.

  As was already mentioned in the
introduction, in the case of the Anderson transition the probability density of the inverse participation ratios $I_q$ was shown to be dependent only on the scaling ratio $z=I_q/I_q^{(t)}$,  with $I_q^{(t)}$ standing for the typical value. Moreover, that ratio is
expected to be power-law distributed: ${\cal P}_q(z)\sim z^{-1-\omega_q}$ \cite{ME,EM}. We may try to combine that fact with the theory
 developed in the present paper to conjecture the typical position of the extreme values (maxima or minima) in a pattern of normalized multifractal probability weights $p_i\sim M^{-\alpha_i}$ for $i=1,\ldots M\sim L^d$ such that $\sum_ip_i=1$. A brief account of such a procedure is as follows. Suppose the mean participation ratios are given by $\overline{I_q}=B(q)\,M^{-\tau_q}$, with a coefficient $B(q)$ of order of unity, and concentrate on those $q$ for which typical and annealed exponents coincide. From it we recover in the usual way the singularity spectrum $f(\alpha)$ by the corresponding Legendre transform: $f(\alpha)\ge 0$ for $\alpha\in[\alpha_{-},\alpha_+]$ and
 further assume $\alpha_{-}>0$ to avoid complications related to the so-called "multifractality freezing" \cite{EM,FRL,F09} which would require a special care. Define $\alpha(q)$ to be a solution of the equation $q=f'(\alpha)$ and denote the mean of the scaling ratio $z=I_q/I_q^{(t)}$ as
 $\overline{z}_q=\int_0^{\infty}{\cal P}_q(z)\, z\, dz$. Further given any function $\phi_q$ of the variable $q$ define a "Lagrange conjugate" function $\phi_*(\alpha)$, by the relation $\phi_*\left(\alpha(q)\right)=\phi_q.$
 Then, by naturally generalizing our earlier consideration of the $1/f$ noise we suggest that the density of exponents defined as $\rho_M(\alpha)=\sum_{i=1}^M\,\delta\left(\frac{\ln{p_i}}{\ln{M}}-\alpha\right)$ should be given asymptotically, in every realization, by the following "improved multifractal Ansatz", cf. (\ref{multiansatzdenhigh}):
\be\label{improved}
\rho_M(\alpha)=\frac{n_*(\alpha)}{\overline{z}_*(\alpha)}B_{*}(\alpha)\sqrt{\frac{\ln{M}|f''(\alpha)|}{2\pi}}\, M^{f(\alpha)}\,\,.
\ee
 Here $n=n_*(\alpha)$ is assumed to be a random coefficient of the order of unity distributed for a given $\alpha$ according to a probability density  ${\cal P}^*_{\alpha}(n)$
defined in terms of the density for the IPR scaling ratio ${\cal P}_q(z)$  via the rule ${\cal P}^*_{\alpha(q)}(n)={\cal P}_q(n)$. Indeed, substituting the Ansatz (\ref{improved}) into the definition $I_q=\int_0^{\infty} M^{-q\alpha}\rho_M(\alpha)\,d\alpha$ and performing the integral
by the Laplace method for $\ln{M}\gg 1$ gives $I_q\approx n_*(\alpha(q))\, \left[\overline{I_q}/\overline{z}_q\right]$, where the random variable $z=n_*(\alpha(q))$ is distributed according to the probability density ${\cal P}_q(z)$. This is precisely what is required, provided we identify
$I_q^{(t)}=\overline{I_q}/\overline{z_q}$. Now we can substitute the same Ansatz to the definition of the counting function $N_<(\alpha)=\int_{-\infty}^{\alpha}\rho_M(\tilde{\alpha})\,d\tilde{\alpha}$ choosing the value of $\alpha$ to the left of the maximum of $f(\alpha)$. This gives asymptotically
$N_{<}(\alpha)=n_*(\alpha) N_t(\alpha)$ where the scale $N_t(\alpha)$ is now given by
\be\label{typicalscale}
N_t(\alpha)=\frac{B_{*}(\alpha)}{\overline{z}_*(\alpha)f'(\alpha)}\sqrt{\frac{|f''(\alpha)|}{2\pi\ln{M}}}\, M^{f(\alpha)}, \quad \alpha_{-}<\alpha<\alpha_0
\ee
and defines the typical value of the counting function for a given $\alpha$. Then, in a typical realization of the multifractal pattern a few "extreme" values among the probability weights $p_i$'s  will be of the order of $p_m=M^{-\alpha_m}$, where $\alpha_m$ is determined from the condition $N_t(\alpha_m)\sim 1$. Clearly, at the leading order $\alpha_m=\alpha_{-}$ given by the left root of $f(\alpha)=0$, and the goal is to extract the subleading term. For doing this properly a crucial observation taken from \cite{ME} is that for $q\to q_{c}=f'(\alpha_-)$ the tail exponent $\omega_{q}$ characterizing the IPR probability density ${\cal P}_q(z)\sim z^{-1-\omega_q}$ should tend to $\omega_{q_{c}}=1$. As the derivative $\frac{d}{dq}\omega_{q}|_{q=q_{c}}$ is generically neither zero nor infinity the mean value
 $\overline{z_q}=\int_0^{\infty}{\cal P}_q(z)\, z\, dz$ will diverge close to $q=q_c$ as $\overline{z_q}\sim (q_c-q)^{-1}$. In turn, as $\overline{z}_*(\alpha)$ is the Lagrange conjugate of
 $\overline{z_q}$ the divergence of the latter implies similar behaviour $\overline{z}_*(\alpha)\sim (\alpha-\alpha_{-})^{-1}$ in the vicinity of $\alpha_{-}$.  At the same time generically  $f'(\alpha), f''(\alpha)$ neither vanish nor diverge at  $\alpha=\alpha_{-}$, and we do not see any reasons to expect that $B_{*}(\alpha)$ vanishes or diverges at this point either. Approximating $f(\alpha_m)\approx f'(\alpha_{-})(\alpha_m-\alpha_{-})$  we arrive at the following equation for the extreme value threshold $\alpha_m$:
\be\label{critextr}
f'(\alpha_{-})(\alpha_m-\alpha_{-})\ln{M}-\frac{1}{2}\ln{\ln{M}}+\ln{(\alpha_m-\alpha_{-})}=0\,.
 \ee
 Solving it for $\ln{M}\gg 1$ to the first non-trivial order beyond $\alpha_m=\alpha_{-}$  gives
 \be\label{last}
 \fl
 \alpha_m\approx \alpha_{-}+\frac{3}{2}\frac{1}{f'(\alpha_{-})}\frac{\ln{\ln{M}}}{\ln{M}} \quad \Rightarrow -\ln{p_m}^{(typ)}\approx \alpha_{-}\ln{M}+ \frac{3}{2}\frac{1}{f'(\alpha_{-})}\ln{\ln{M}}
 \ee
which constitutes our main prediction for the typical position of the threshold of extreme values in disorder-induced multifractals.
In particular, the value of the absolute maximum $p_{max}$ will be such that $y=\ln{p_{max}}-\ln{p_m^{(typ)}}$ is a random variable of the order of unity.

\section{Discussion and Conclusion}
In conclusion, we have studied, both analytically and numerically, the strongly fluctuating multifractal pattern associated with high values of the periodic ideal $1/f$ noise. In particular, we concentrated on the signal level comparable with the typical maximum value of the $1/f$ noise.  The exploitation of the thermodynamic formalism allowed us to translate the distribution of the partition function found in the previous studies \cite{FB,FLDR2009} to a similar distribution for the counting functions of exceedances of such a high level. The power-law forward tail of the latter distribution was shown to give rise to a parametric difference between the mean and the typical value of the counting function when the position of the high level approaches threshold $x_m$ of extreme values. Such a mechanism which can be traced back  to the logarithmic correlations inherent in the $1/f$ noise allowed us to explain the universal coefficient in front of the subleading term in the position of the threshold $x_m$.

We have also performed a direct numerical simulations of the $1/f$ signal and calculated numerically the lowest two moments of the partition function.
This served to demonstrate that for the samples of $M\sim 10^6$ points the numerics follows $M=\infty$ results rather faithfully.
Performing the same check for the counting function moments however showed that truly asymptotic results can be never achieved even with very moderate accuracy due to prohibitively slow convergence. Instead, even for $M$ as big as $M\sim 10^9$ one has still to use a more elaborate finite-$M$ formulas which we derived in the present paper for that goal. This lesson may prove important in view of the growing interest in numerical simulations of
related systems arising in the framework of the Random Matrix Theory and the Riemann zeta function along the critical line \cite{FHK}.

Finally, by comparing the results obtained for $1/f$ noises in our paper with those known to hold for multifractal patterns of wave-function intensity at the points of Anderson transitions \cite{EM,ME} we propose a quite general formula (\ref{last}) for the position of extreme values in generic disorder-generated multifractal patterns with non-parabolic singularity spectra. We hope that such prediction can be checked against the accurate numerical data in random multifractals of various origin, and will generate further interest in statistics of
high and extreme values in such multifractal patterns. We leave this issue as well as a related, but much more difficult question about actual statistics of the counting function in the region of extreme states for future investigations. For $1/f$ noise the latter should involve understanding of how the so-called "freezing phenomena" known to have profound influence on the behaviour of the partition function $Z_q$ with $|q|>1$ are reflected in the thermodynamic formalism correspondence between $Z_q$ and the counting function. Note that the freezing mechanism suggested in \cite{CLD,FB,FLDR2009} predicts for $|q|>1$ the tail behaviour for the distribution of $Z=Z_q$ to be ${\cal P}_{|q|>1}(Z)\sim  Z^{-\left(1+\frac{1}{|q|}\right)}\ln{Z}$. It is based on considering properties of the generating function $g_q(y)=\overline{\exp\left\{-e^{-qy}Z_q/Z_e(q)\right\}}$ which in the limit $M\to \infty$ is conjectured to stay $q-$independent (i.e. "frozen") to the value $g_{|q|=1}(y)$ for all $|q|>q_c=1$.
The factor $\ln{Z}$ in ${\cal P}_{|q|>1}(Z)$ plays a prominent role and is believed to be a universal feature within the class of Gaussian logarithmically correlated fields. Note that such factor will be precisely absent for i.i.d. case of the standard REM model. To that end
 let us mention that in the context of the Anderson Localization a certain use of the thermodynamic formalism for the IPR's combined with  a clever heuristic power counting \cite{ME,EM} lead to predicting the probability density for $I=I_q$ with $q>q_c=f'(\alpha_{-})$ of the form ${\cal P}_{q>q_c}(I)\sim  I^{-\left(1+\frac{q_c}{q}\right)}$. It is most natural to suspect that the logarithmic factor $\ln{I}$ should be present in the above formula for ${\cal P}_{q>q_c}(I)$ as well, and the accuracy of the power counting procedure used in \cite{ME,EM} was simply not enough to account for it. Closely related questions are whether the generating function $\tilde{g}_q(y)=\overline{\exp\left\{-e^{-qy}I_q/I_q^{(t)}\right\}}$
will be actually $q-$independent for $q>q_c=f'(\alpha_{-})$ for general non-parabolic multifractals and whether the probability density of the logarithm of the (appropriately shifted) absolute maximum $y=\ln{p_m^{(typ)}}-\ln{p_{max}}$, with $p_m^{(typ)}$ given by (\ref{last}), will show a characteristic non-Gumbel tail $|y|e^{-|y|}, y\to -\infty$ \cite{CLD,FB,FLDR2009} as our extended analogy would suggest. All these intriguing issues certainly deserve further investigation, both numerically and analytically.

\vspace{1cm}

{\bf Acknowledgements.} YVF is grateful to Ilya Gruzberg and Vincent Vargas for their interest in the work and useful comments.
He also acknowledges LPTENS for the financial support and hospitality during early stages of the project, and the programme "Disordered Quantum Systems" at the Institut Henri Poincare for a support during the completion stage.
 The research at Queen Mary University of London was supported by EPSRC grant EP/J002763/1 ``Insights into Disordered Landscapes via Random Matrix Theory and Statistical Mechanics''.

\vspace{1cm}

\appendix{\bf Appendix A: Variance of the counting function for Gaussian sequences.}

\vspace{1cm}
Suppose we have a sequence of $M$ correlated Gaussian-distributed variables $V_1,\ldots,V_M$ characterized by the common variances
$\overline{V_i^2}=c_{0}$ and covariances  $\overline{V_iV_j}=c_{ij},\, i\ne j$. Define
 \be\label{gendefinitiondelta}
 \Delta_M(a)=\sum_{i\ne j} \left[\overline{\theta(V_i-a) \theta(V_j-a)}- \overline{\theta(V_i-a)}\cdot \overline{ \theta(V_j-a)} \right]
\ee
 Our goal is to show the validity of the following expression
\be\label{exdelta}
\Delta_M(a)=\frac{1}{\pi}\exp{\left(-\frac{a^2}{2c_0}\right)}\sum_{i\ne j}^{M} \int_{h_{ij}}^1\frac{d\tau}{1+\tau^2}e^{-\frac{a^2}{2c_0}\tau^2},
\quad h_{ij}=\sqrt{\frac{c_0-c_{ij}}{c_0+c_{ij}}}
\ee
For proving it we need the following \\
{\bf Proposition}. Suppose $V_1,V_2$ are two Gaussian-distributed variables characterized by the common variances
$\overline{V_1^2}=\overline{V_1^2}=c_{0}$ and the covariance  $\overline{V_1V_2}=c$. Then for any two functions $f_1(V)$ and $f_2(V)$
with finite means $\overline{f_{1,2}(V)}$  holds the identity
\be\label{gauident}
\frac{\partial}{\partial c}\left[\overline{f_1(V_1)\, f_2(V_2)}\right]=\overline{f'_1(V_1)\, f'_2(V_2)}
\ee
where $f'\equiv \frac{df}{dV}$.
To verify the proposition we introduce the vector ${\bf v}=\left(\begin{array}{c}V_1\\V_2\end{array}\right)$, denote $\hat{C}=\left(\begin{array}{cc} c_0 & c\\ c & c_0 \end{array}\right)$ and write the joint probability density of $V_1$ and $V_2$ as
\be
{\cal P}({\bf v})=\frac{e^{-{\bf v}^T \hat{C}^{-1}\bf{v} }}{2\pi\sqrt{det{\hat{C}}}}=\int_{-\infty}^{\infty}\int_{-\infty}^{\infty}
 e^{-\frac{1}{2}\phi^T \hat{C} \phi+i(V_1\phi_1+V_2\phi_2)}\frac{d\phi_1d\phi_2}{2\pi}
\ee
from which it is immediately clear that
\be
\frac{\partial}{\partial c}{\cal P}({\bf v})=\frac{\partial^2}{\partial V_1\partial V_2}{\cal P}({\bf v})
\ee
This implies:
 \be\label{gauident1}
 \fl
\frac{\partial}{\partial c}\left[\overline{f_1(V)\, f_2(V)}\right]=\int_{-\infty}^{\infty}
\int_{-\infty}^{\infty} f_1(V_1)\, f_2(V_2)\frac{\partial^2}{\partial V_1\partial V_2}{\cal P}({\bf v})dV_1dV_2=
\overline{f'_1(V)\, f'_2(V)}
\ee
where the last equality follows after integration by parts.

Now applying the proposition to the particular case $f_1(V)=f_2(V)=\theta(V-a)$ gives in view of $\frac{\partial}{\partial V}\theta(V-a)=\delta(V-a)$
the relation
\be
\fl
\frac{\partial}{\partial c}\left[\overline{\theta(V_1-a)\, \theta(V_2-a) }\right]={\cal P}(V_1=a,V_2=a)=\frac{1}{2\pi\sqrt{c_0^2-c^2}}\,e^{-\frac{a^2}{c_0+c}}
\ee
and since $\overline{\theta(V_1-a)}=\overline{\, \theta(V_2-a) }$  is independent of $c$ the integration shows that
\be
\fl
D_{12}=\overline{\theta(V_1-a)\, \theta(V_2-a) }-\overline{\theta(V_1-a)}\cdot \overline{\, \theta(V_2-a) }=\frac{2}{\pi}\int_{0}^{c}\frac{du}{\sqrt{c_0^2-u^2}}\,e^{-\frac{a^2}{c_0+u}}
\ee
 Finally, introducing the variable $\tau=\sqrt{\frac{c_0-u}{c_0+u}}$ we convert the above expression to the form
\be\label{deltatermfin}
\fl
D_{12}=
\frac{4}{\pi}\exp{\left(-\frac{a^2}{2c_0}\right)} \int_{h}^1\frac{d\tau}{1+\tau^2}e^{-\frac{a^2}{2c_0}\tau^2},
\quad h=\sqrt{\frac{c_0-c}{c_0+c}}
\ee
which after applying to each pair of $V_i,V_j$ in (\ref{gendefinitiondelta}) immediately implies (\ref{exdelta}).

\vspace{1cm}

\appendix{\bf Appendix B: Large-$M$ asymptotic of the $\delta_M^{(n)}(x)$ for $\sqrt{2}<x<2$.}\\

\vspace{0.5cm}

Our goal is to extract the large-$M$ asymptotic behaviour of the integral featuring in (\ref{approxdelta1}), that is:
\be\label{approxdelta1}
\fl
J_M(x)=\int_{\frac{\pi}{M}}^{\frac{\pi}{2}}d\theta \int_{h_{\theta}}^1\frac{d\tau}{1+\tau^2}e^{-\ln{M}\,x^2\tau^2/4}, \quad h_{\theta}=\sqrt{\frac{\ln{M}+\ln{|2\sin{\theta}|}}
{\ln{M}-\ln{|2\sin{\theta}|}}}
\ee
For $x>\sqrt{2}$ we anticipate that the main contribution for $M\to \infty$ comes from $\theta\to 0$ and rescale the integration variable
as $\theta=\pi M^{-u}, u\in (0,1)$. Such a rescaling allows us to replace $h(\theta)\to h(u)=\sqrt{\frac{1-u}{1+u}}$ so that we have
\be\label{approxdelta2}
J_{M\gg 1}(x)\approx -\pi\int_{0}^{1}d\left(e^{-\ln{M}u}\right)
 \int_{h(u)}^1\frac{d\tau}{1+\tau^2}e^{-\ln{M}\,x^2\tau^2/4}
\ee
 \[
 =\pi M^{-1} \int_0^1\frac{d\tau}{1+\tau^2}e^{-\ln{M}\,x^2\tau^2/4}- \frac{\pi}{2}\int_0^1\, du \, h'(u) (1+u)\,
 e^{-\ln{M}\left(u+\frac{x^2}{4}\frac{1-u}{1+u}\right)}
 \]
 For $\ln{M}\gg 1$ the first integral is dominated by the lower limit and yields the leading order contribution $J^{(I)}_{M\gg 1}(x)\approx \frac{\pi}{M\,x}\sqrt{\frac{2\pi}{\ln{M}}}$. Second integral is dominated by the vicinity of the minimum of the function
 ${\cal L}(u)=u+\frac{x^2}{4}\frac{1-u}{1+u}$  achieved at $u_*=\frac{x}{\sqrt{2}}-1$. For $\sqrt{2}<x<2\sqrt{2}$ we have $u_*\in(0,1)$
 so we can apply the standard Laplace method. Using ${\cal L}(u_*)=-1+\sqrt{2}x-x^2/4$ and $\frac{d^2}{du^2}{\cal L}(u=u_*)=2\sqrt{2}/x$ we find the leading order
contribution given by:
\be
J^{(II)}_{M\gg 1}(x)\approx \sqrt{\frac{\pi^{3}}{2\sqrt{2}\ln{M}(2\sqrt{2}-x)}}\,M^{1-\sqrt{2}x+\frac{x^{2}}{{4}}}
\ee
Finally, using asymptotic formula $\mbox{Erfc}\left(\frac{x}{2}\sqrt{\ln{M}}\right)\approx \frac{2M^{-\frac{x^2}{4}}}{x\sqrt{\pi\ln{M}}}$
we arrive at the expression (\ref{approx4}).

\end{document}